\documentclass{aa}             

\usepackage{natbib}
\bibpunct{(}{)}{;}{a}{}{,} 

\input epsf

\usepackage{graphicx}
\usepackage{txfonts}

\begin{document}

\title{
Age and helium content of the open cluster NGC\,~6791\\
from multiple eclipsing binary members
\thanks{Based on observations carried out at the 
Nordic Optical Telescope at La Palma and ESO's VLT/UVES
ESO, Paranal, Chile (75.D-0206A, 77.D-0827A, 81.D-0091).}
\fnmsep\thanks{Tables 5 and 6 are only available in electronic form
at the CDS via anonymous ftp to cdsarc.u-strasbg.fr
(130.79.128.5) or via http://cdsweb.u-strasbg.fr/cgi-bin/qcat?J/A+A/}
} 
\subtitle{II. age dependencies and new insights}

\author{
K. Brogaard       \inst{1,2}
\and D. A. VandenBerg	\inst{1}
\and H. Bruntt    \inst{2} 
\and F. Grundahl  \inst{2} 
\and S. Frandsen  \inst{2}
\and L. R. Bedin	\inst{10,3}
\and A. P. Milone	\inst{4}
\and A. Dotter  	\inst{3}
\and G. A. Feiden	\inst{5}
\and P. B. Stetson      \inst{6}
\and E. Sandquist	\inst{7}
\and A. Miglio		\inst{8}
\and D. Stello		\inst{9}
\and J. Jessen-Hansen	\inst{2}
}

\offprints{K. Brogaard, e-mail: kfb@phys.au.dk}

\institute{
Department of Physics and Astronomy,
University of Victoria,
P.O. Box 3055, Victoria, B.C., V8W 3P6, Canada
\and
Department of Physics and Astronomy,
Aarhus University, 
Ny Munkegade, DK-8000 Aarhus C, Denmark
\and
Space Telescope Science Institute, 
3700 San Martin Drive, Baltimore, MD 21218, USA
\and
Instituto de Astrofisica de Canarias, 
E-38200 La Laguna, Tenerife, 
Canary Islands, Spain
\and
Department of Physics and Astronomy,
Dartmouth College,
6127 Wilder Laboratory, Hanover, NH 03755, USA
\and
Dominian Astrophysical Observatory, Herzberg Institute of Astrophysics, National Research Council, 5071 West Saanich Road, Victoria, B.C., V9E 2E7, Canada
\and
Department of Astronomy, San Diego State University, San Diego, CA 92182, USA
\and
School of Physics and Astronomy, University of Birmingham, Edgbaston, Birminghma B15 2TT, United Kingdom
\and
Sydney Institute for Astronomy (SIfA), School of Physics, University of Sydney, NSW 2006, Australia
\and
INAF - Osservatorio Astronomico di Padova, vicolo dell'Osservatorio 5, 35122 Padova, Italy
}

\date{Received xx XX 2012 / Accepted xx XX 2012, version 16.05.2012}
 
\titlerunning{Eclipsing binaries in the open cluster NGC\,6791}
\authorrunning{K. Brogaard et al.}

\abstract
{Models of stellar structure and evolution can be constrained by measuring accurate parameters of
 detached eclipsing binaries in open clusters. Multiple binary stars provide the means to determine helium abundances in these old stellar systems, and in turn, to improve estimates of their age.
}
{In the first paper of this series, we demonstrated how measurements of multiple eclipsing binaries in the old open cluster NGC\,6791 sets tighter constraints on the properties of stellar models than has previously been possible, thereby potentially improving both the accuracy and precision of the cluster age. Here we add additional constraints and perform an extensive model comparison to determine the best estimates of the cluster age and helium content, employing as many observational constraints as possible.
}
{We improve our photometry and correct empirically for differential reddening effects. We then perform an extensive comparison of the new colour-magnitude diagrams (CMDs) and eclipsing binary measurements to Victoria and DSEP isochrones in order to estimate cluster parameters. We also reanalyse a spectrum of the star 2-17 to improve [Fe/H] constraints.
}
{We find a best estimate of the age of $\sim$\,8.3 Gyr for NGC\,6791 while demonstrating that remaining age uncertainty is dominated by uncertainties in the CNO abundances. The helium mass fraction is well constrained at $Y=0.30\pm0.01$ resulting in $\Delta Y/\Delta Z \sim$\,1.4 assuming that such a relation exists. During the analysis we firmly identify blue straggler stars, including the star 2-17, and find indications for the presence of their evolved counterparts. Our analysis supports the RGB mass-loss found from asteroseismology and we determine precisely the absolute mass of stars on the lower RGB, $M_{\rm RGB} = 1.15\pm0.02 M_{\sun}$. This will be an important consistency check for the detailed asteroseismology of cluster stars.
}
{Using multiple, detached eclipsing binaries for determining stellar cluster ages, it is now possible to constrain parameters of stellar models, notably the helium content, which were previously out of reach. By observing a suitable number of detached eclipsing binaries in several open clusters, 
 it will be possible to calibrate the age--scale and the helium enrichment parameter $\Delta\,Y/\Delta\,Z$, and provide firm constraints that stellar models must reproduce.
}
\maketitle

\section{Introduction}
\label{sec:intro}

The open cluster NGC~6791 is one of the oldest and, at the same time, the most metal-rich open clusters known (\citealt{Origlia06,Carretta07,attm07, Brogaard11}; hereafter paper I), and is therefore very important for understanding chemical evolution and metal-rich stars. In addition, it is a well-populated open cluster, with a (statistically) significant number of stars in all stages of evolution from the main sequence to the white dwarfs \citep{King05, Bedin05, Bedin08}, as well as with numerous variable stars \citep{Bruntt03,Mochejska02,deMarchi07}. 

Despite a number of studies that have presented well-calibrated photometry and high-precision colour-magnitude diagrams (CMDs) (\citealt{King05,SBG03}), the age of NGC~6791 has remained very uncertain until recently because of correlated uncertainties in distance, reddening, and metallicity.

 It is widely appreciated that detached eclipsing binaries 
 offer the possibility of determining accurate (and precise) masses and radii for the 
 system components, nearly independent of model assumptions \citep{Andersen91,Torres10}. If the binary resides in a star cluster, and one or both of its components are close to the turn-off, it is possible to put tight constraints on the age of the system by comparing the position of the primary and secondary in a mass--radius (MR) diagram with theoretical isochrones. For stellar clusters, such an analysis has some
 significant advantages: the determination of the masses and radii is 
 independent of the usual uncertainties such as reddening and distance. Furthermore, since the 
 comparison to models is carried out in the MR diagram, one avoids
 the difficult process of transforming the effective temperatures and 
 luminosities of the models to observed colours and magnitudes. Thus, determining cluster ages in the MR diagram allows a direct comparison between 
 observations and theory.
 
\cite{Grundahl08} showed that, by applying this method on the eclipsing binary V20, a cluster member, they could determine a precise cluster age with an uncertainty of only $\pm0.3$ Gyr for a given stellar model. However, they were unable to determine which of the models (if any) to trust, because the difference in the predicted age, depending on which specific stellar model they adopted, was about four times greater than their measurement precision. 

In Paper I, we undertook an analysis of three detached eclipsing 
 binary systems, V18, V20 and V80, in NGC\,6791, and determined accurate masses and radii for the components of the first two of these systems. We also measured spectroscopic $T_{\rm eff}$\ and $[\mathrm{Fe/H}]$\ values 
 from disentangled spectra of the binary stars and a few single stars. This was used, together with the cluster CMD, in a first demonstration of how measurements of multiple eclipsing binaries in a star cluster constrain stellar models and cluster parameters, like age and helium content, better than previously possible. Here we add additional constraints from asteroseismology and do an extensive model comparison to determine a best-estimate of the helium content and cluster age while quantifying remaining age dependencies. As it turns out, this analysis also results in new insights on evolved cluster stars.

\section{Observational data}
\label{sec:data}

For the present analysis we employ the observational data and measurements derived in paper I with slight modifications and additions as described in this section. 

\begin{table*}   
\begin{center}
\caption[]{\label{tab:absdim}
Astrophysical data for the eclipsing binaries V18 and V20.  
}
\begin{tabular}{lrrrr} \hline    
\hline    
\noalign{\smallskip}    
                     &    \multicolumn{2}{c}{V18}       &    \multicolumn{2}{c}{V20}  \\ 
\noalign{\smallskip}    
                     &  \multicolumn{1}{c}{Primary}   & \multicolumn{1}{c}{Secondary} & \multicolumn{1}{c}{Primary} &    \multicolumn{1}{c}{Secondary}     \\ 
\noalign{\smallskip}    
\hline    
\noalign{\smallskip}    
$M/M_{\sun}$     &$0.9955 \pm 0.0033$ &$0.9293 \pm 0.0032$ &$1.0868 \pm 0.0039$ &$0.8276 \pm 0.0022$ 
\\ 
$R/R_{\sun}$     &$1.1011 \pm 0.0068$ &$0.9708 \pm 0.0089$ &$1.397 \pm 0.013$  &$0.7813 \pm 0.0053$ 
\\ 
$\log g$ (cgs)   &$4.3524 \pm 0.0053$ &$4.4319 \pm 0.0080$ &$4.1840 \pm 0.0078$ & $4.5698 \pm 0.0059$ 
\\ 
$T_{\mbox{\scriptsize eff}}\,$ (K)\tablefootmark{a}  &  $5600 \pm 95$ &   $5430\pm 125$ &   $5645 \pm 95$ & $\mathbf{4860 \pm 125}$\tablefootmark{a} \\

$V$				    & $\mathbf{18.256\pm0.020}$	& $\mathbf{18.748\pm0.020}$	& $\mathbf{17.702\pm0.020}$ & $\mathbf{19.886 \pm 0.020}$ \\
$BC_V$				    & $\mathbf{-0.084\pm0.018}$	& $\mathbf{-0.123\pm0.034}$	& $\mathbf{-0.074\pm0.018}$ & $\mathbf{-0.344}$\tablefootmark{a} \\
$M_V$				    & $4.76\pm0.12$     & $5.20\pm0.16$         & $\mathbf{4.20\pm0.09}$ & $\mathbf{6.38\pm0.15}$\tablefootmark{a}\\
$(m-M)_V$			    & $\mathbf{13.50\pm0.12}$	& $\mathbf{13.55\pm0.16}$	& $\mathbf{13.51\pm0.09}$ & $\mathbf{13.51\pm0.09}$\tablefootmark{a}\\
\noalign{\smallskip}            
\hline
\end{tabular}
\end{center}            
\tablefoot{Measurements from paper I, slightly updated using new photometry and a new approach for V20s. Values that are different from, or were not included in, paper I are in boldface.
\\
\tablefoottext{a}{$T_{\rm eff}$\ values are spectroscopic measurements, except for $T_{\rm eff}$\ of V20s, which is calculated assuming the same distance modulus as for V20p.}
}
\end{table*}                   

\subsection{Photometry \& binary measurements}
\label{sec:phot}

In paper I we used the photometry of \cite{SBG03} for some of the binary measurements, while the CMDs were generated from a newer unpublished reduction of the same data performed by P. B. Stetson. At that time we thought that differences would be negligible and of random character. However, we found later that there are systematic differences between these data sets, due to issues with the photometric zero-points, as described in \cite{Stetson05} and corrected in the new reduction. 

We have therefore redone those binary measurements which depend on the photometry, such that our analysis is now consistently making use of the new photometric reduction. 

Only precise measurements are useful and we therefore concentrate on the measurements of the eclipsing binaries V18 and V20, leaving out the less precise measurements of V80 from paper I. In Table~\ref{tab:absdim} we show our measurements for these systems from paper I with the slightly revised values of $V, M_V$ and $(m-M)_V$. As seen, changes to the magnitudes are small and the error-weighted mean apparent distance modulus remains at $(m-M)_V=13.51 \pm 0.06$, the value found in paper I.

Since the following analysis will show that the reddening and colour-temperature relations are rather uncertain, we calculated the $T_{\rm eff}$ for the V20 secondary component, which does not have a spectroscopic $T_{\rm eff}$, under the reasonable assumption that it has the same distance modulus as the V20 primary component. This avoids the dependence on the reddening and colour-temperature relations which was present for this value in paper I. We have also chosen not to use a reddening measurement as a direct constraint for the same reasons, since our only precise reddening measurement in paper I relied on colour-temperature relations.

\subsection{Differential reddening correction}
\label{sec:difred}

Several authors have noted spreads in the CMD sequence of NGC\,6791 that is larger than can be explained by photometric errors alone (\citealt{Bedin08, Twarog11, Platais11}). The most plausible explanation for this is differential reddening effects, especially since offsets from a mean fiducial are correlated with position on the sky (\citealt{Twarog11, Platais11}).

To correct for the differential reddening we used an empirical method described
in detail in \cite{Milone12}.
 Briefly, we define the fiducial main
sequence for the cluster and estimate for each star how the
observed stars in its vicinity may
systematically lie to the red or the blue of the fiducial sequence;
this systematic colour offset is indicative of the local differential
reddening.

\begin{figure*}
\epsfxsize=180mm
\epsfbox{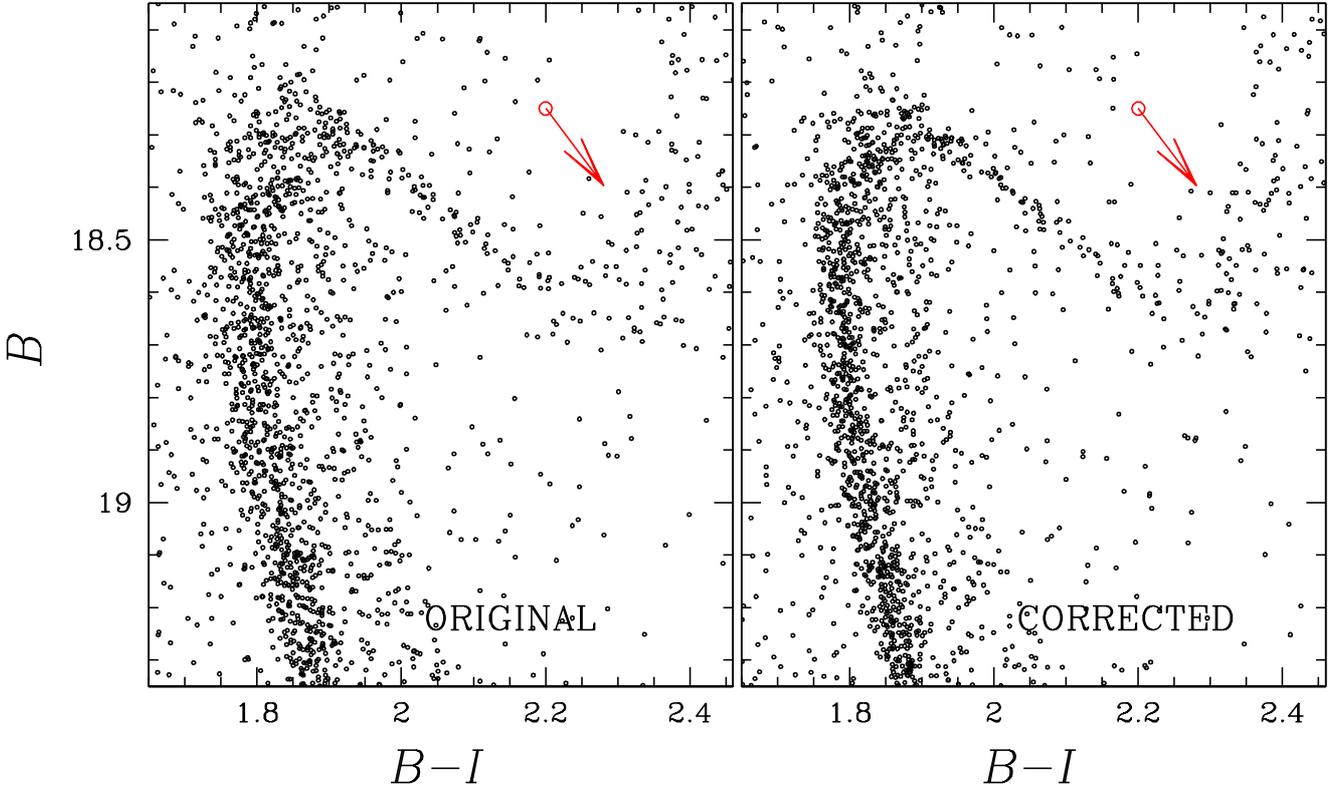}
\caption[]{\label{fig:difred}
The $B,B-I$ CMD of NCG~6791. Left: as observed. Right: after empirical correction for differential reddening.
}
\end{figure*}

In Fig.~\ref{fig:difred}, we compare the original (left panel) and the
corrected CMD (right panel) of NGC\,6791.
It should be noted how many features of the CMD become
narrower and more clearly defined after the correction has been
applied, confirming that most of the effects of
differential reddening have been removed.  The improvement of the CMD
is particularly evident for the sub-giant branch (SGB) and the lower red giant branch (RGB); the tightness along the reddening line of these stars indicate that the spread we see along the SGB is mainly due to differential
reddening.

\begin{figure}
\epsfxsize=95mm
\epsfbox{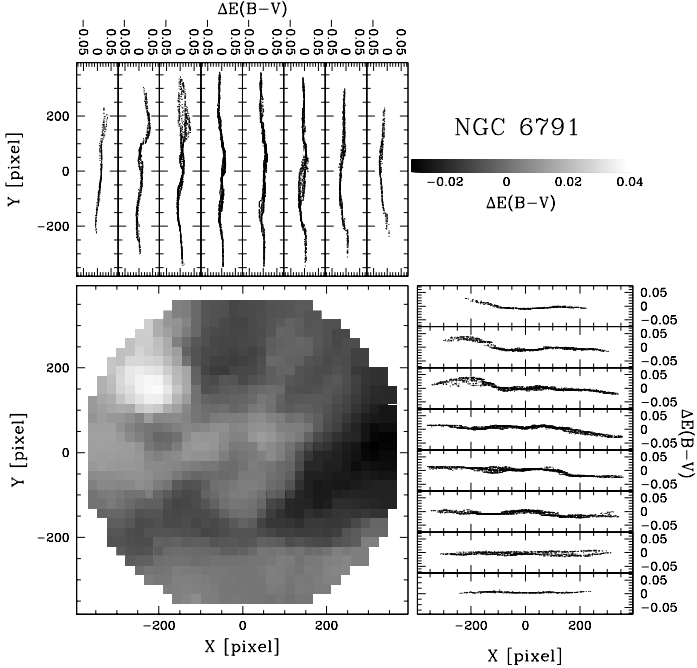}
\caption[]{\label{fig:difredmap}
Differential reddening map of the central part of NGC\,6791. (x,y) coordinates are in arcseconds relative to 19:20:53.22 +37:46:18.8 (2000.0).
}
\end{figure}

From star-to-star comparisons of the original and the corrected {\it V}
magnitudes we can estimate star-to-star variations in $E(B-V)$
and derive the reddening map in the direction of NGC\,6791.
This is shown in Fig.~\ref{fig:difredmap}, where we have divided the field of view
into 8 horizontal slices and 8 vertical slices and plot
$\Delta\,E(B-V)$ as a function of the Y (upper panels) and X
coordinate (right panels).
We have also divided the whole field of view into 64$\times$64 boxes
and calculated the average $\Delta~E(B-V)$ within each of them.
The resultant reddening map is shown in the lower-left~panel
where each box is represented as a gray square. The levels of gray
are indicative of the amount of differential reddening as shown in the
upper-right panel. We cannot know for sure that this map shows only the effects of differential reddening, since systematic effects from the instrument and/or the reduction procedure may also be present in the map. However, we repeated the procedure using {\it HST} photometry for a smaller field within our FOV, and found a strong correlation with the ground-based results, indicating that the main effect is non-local.

The photometry, with and without the differential reddening correction, is available in Tables 5 and 6 on CDS.

\subsection{Additional model constraints}
\label{sec:debs}

To put additional constraints on the stellar models we include in our model comparisons, along with the CMDs and binary measurements, the RGB mass-loss in the form of the mean mass difference $\Delta M=0.09\pm0.03\mathrm{(random)}\pm0.04\mathrm{(systematic)}$ $M_{\sun}$ between the lower RGB and the red clump (RC), as measured by \cite{Miglio12} using asteroseismology. We employ only the mass difference because the seismic absolute masses are too uncertain and potentially dominated by systematic errors to be useful at this point. For our model comparisons we therefore mark the region on the zero-age horizontal branch (ZAHB) with masses from that of the RGB tip down to a mass lower by twice the mean mass-loss (to allow for a mass-loss dispersion).
We also make use of our metallicity measurement from paper\,I, $\mathrm{[Fe/H]}=+0.29\pm0.03\mathrm{(random)}\pm0.08\mathrm{(systematic)}$ for the turn-off and early SGB region, in combination with an additional [Fe/H] measurement described in Sect.~\ref{sec:bhb}.

\section{Comparison to theoretical models}
\label{sec:comp}

\subsection{Stellar models}
\label{sec:models}

\begin{figure}
\epsfxsize=80mm
\epsfbox{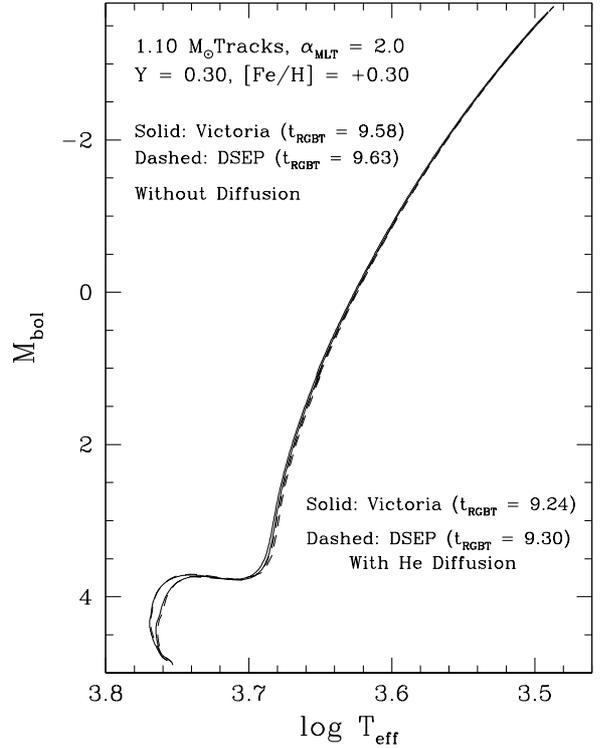}
\caption[]{\label{fig:uvicdsepp03}
Comparison between Victoria and DSEP stellar evolutionary tracks for masses
and chemical abundances (as indicated) that are relevant to an investigation
of NGC~6791 and the binary stars V18 and V20. Note that there are four tracks in the figure, with those without diffusion being hottest in the turn-off region.}
\end{figure}

The stellar models that are used in the analysis of the
binary stars and the cluster CMDs, were
generated using a significantly updated version of the University of Victoria
evolutionary code (\citealt*{VandenBerg06}; and references
provided therein) with updates summarized in paper I. For a much more detailed description, we refer to a
forthcoming study which examines the impact of varying the abundances of
individual heavy elements (D.~VandenBerg et al. 2012, submitted). 
Since the settling of heavy elements is not (yet) considered in the Victoria models, we include also a few models calculated using the DSEP code (\citealt{Feiden11} and references therein) which includes heavy element diffusion using the description of \cite{Thoul94} and turbulent mixing as described by \cite{Richard05}. As seen in Fig.~\ref{fig:uvicdsepp03}, the DSEP and Victoria codes produce nearly identical results for identical input physics.  We can therefore be confident that it is mainly the effects of heavy-element diffusion we see when comparing Victoria and DSEP models with identical composition but including heavy-element diffusion in the latter.
All models are calibrated in order to satisfy the solar constraints and employ the surface boundary conditions of \cite{Krishna66}. This was shown by \cite{VandenBerg08} to give a better match to the RGB of the solar metallicity open cluster M67 than when surface boundary conditions from 1D model atmospheres were used. Solar calibration values for each model are given in Table~\ref{tab:calib}.

\begin{table}   
\begin{center}
\caption[]{\label{tab:calib}
Solar calibration values obtained for each stellar model.  
}
\begin{tabular}{lllccc} \hline    
\hline    
\noalign{\smallskip}    
code & element mix & diffusion & $Z_i$\tablefootmark{d} & $Y_i$\tablefootmark{d} & $\alpha _{\rm MLT}$ \\
\hline    
\noalign{\smallskip}    
Victoria & AGS09\tablefootmark{a} & $Y,Tx$\tablefootmark{b} & 0.01323 & 0.2553 & 2.005 \\
Victoria & GS98\tablefootmark{a} & $Y,Tx$\tablefootmark{b} & 0.01652 & 0.2661 & 2.007 \\
DSEP & GS98\tablefootmark{a} & $Y,Z$ & 0.01875 & 0.2743 & 2.139 \\
DSEP & GS98\tablefootmark{a} & $Y,Z,T60$\tablefootmark{c} & 0.01875 & 0.2743 & 2.139 \\
DSEP & GS98\tablefootmark{a} & $Y,Z,T65$\tablefootmark{c} & 0.01749 & 0.2685 & 2.087 \\

\noalign{\smallskip}            
\hline
\end{tabular}
\end{center}            
\tablefoot{
Calibrations were done using surface boundary conditions according 
to \cite{Krishna66}.\\
\tablefoottext{a}{AGS09: \cite{Asplund09}, GS98: \cite{Grevesse98}}
\tablefoottext{b}{turbulence is included, details in Vandenberg et al. (2012, submitted).}
\tablefoottext{c}{turbulence is included, details in \cite{Feiden11}.}
\tablefoottext{d}{$Z_i$ and $Y_i$ refer to the initial value of metal and helium mass fractions.}
}
\end{table}

\subsection{General comparison procedure}
\label{sec:proc}

The best starting point for an analysis of binary stars is arguably the
mass-radius diagram because it provides the most direct comparison
between theory and observations that can be made, being independent of
uncertainties in distance, reddening, and colour--temperature transformations.

For the initial analysis in paper I, we scaled the
relative abundances of the heavy elements recently derived for the Sun by
\cite{Asplund09} and allowed a range in $[\mathrm{Fe/H}]$\ from $+0.2$ to $+0.4$ with the helium mass-fraction $Y$ as a free parameter. Within those parameters we were not able to find a model which matched the observations of the binaries and the CMD as well as we had hoped/expected and agreement in the mass-radius diagram was not better than at the 2-$3\,\sigma$ level. Here we perform a detailed analysis, allowing also for the alternative abundance pattern of \cite{Grevesse98} as well as a variation in the CNO abundances.

Interestingly, even within this expanded parameter space we are still unable to find a very good overall match to all observations; models that optimise the match in the mass-radius diagram of the binaries are too hot to fit the $T_{\rm eff}$s and too young to fit the observed CMDs. This could be a sign that our observations have reached a precision level which cannot be reproduced by current stellar models. However, it is also likely that our precision measurements have entered a regime where systematic effects are dominating instead of random errors. In order to avoid overinterpretation we therefore accept models which match the binary mass and radius measurements within $3\,\sigma$, and their $T_{\rm eff}$s within $1\,\sigma$. We chose the tighter constraint on $T_{\rm eff}$ because $T_{\rm eff}$ is highly correlated with [Fe/H] and the distance modulus. Hence, if the measured $T_{\rm eff}$s are wrong by more than $1\,\sigma$ due to random errors, so are the estimated metallicity and distance modulus, and the model should then match a whole other constraint set. Tests showed that such adjustments only makes the fit worse. The $T_{\rm eff}$ uncertainties include a systematic contribution of 70 K (paper I) and we consider it unlikely that our spectroscopic temperature scale is different from the real one by more than the total $1\,\sigma$ uncertainties which are $\sim$\,100 K.

Since we found in paper I that optimising the fit in the mass-radius diagram does not lead to the best overall fit, the general procedure adopted to find the best parameters of a given model was as follows: First, the age of a model with given parameters was adjusted to make an isochrone match the two most massive, and therefore most age-sensitive, binary components in the mass-radius diagram. The corresponding isochrone in the CMDs would in general not match the SGB $V$-magnitude at the observed distance modulus, which we held fixed at the value measured for the binaries (see Sec.~\ref{sec:precision} for the age uncertainty introduced by this choice); if the predicted SGB was too bright we increased [Fe/H] or decreased $Y$ in the model until the corresponding isochrone matched both the most massive components in the mass-radius diagram and the SGB $V$-magnitude. We emphasize that this procedure is only possible because we have determined the spectroscopic $T_{\rm eff}$s of the binary components and thus the distance modulus in a way which is independent of the CMD and the reddening. 

$E(V-I)/E(B-V)=1.244$ was chosen for all plots. This corresponds to an extreme lower limit found in the literature by using the reddening law of \cite{Cardelli89}, and assuming $R_V=3.1$ and effective wavelengths from \cite{McCall04} for a Vega-like star. We choose this extreme limit instead of a more reasonable value in order to show that even allowing an extreme limit, we still have difficulties obtaining consistency among CMDs in different colour-planes. The reddening value for each isochrone was chosen to maximize consistency between colour-planes, with least weight given to the $B-V$ colour, since that turned out to be the most problematic one.

\subsection{Age dependencies}
\label{sec:age}

\begin{figure*}
\epsfxsize=190mm
\epsfbox{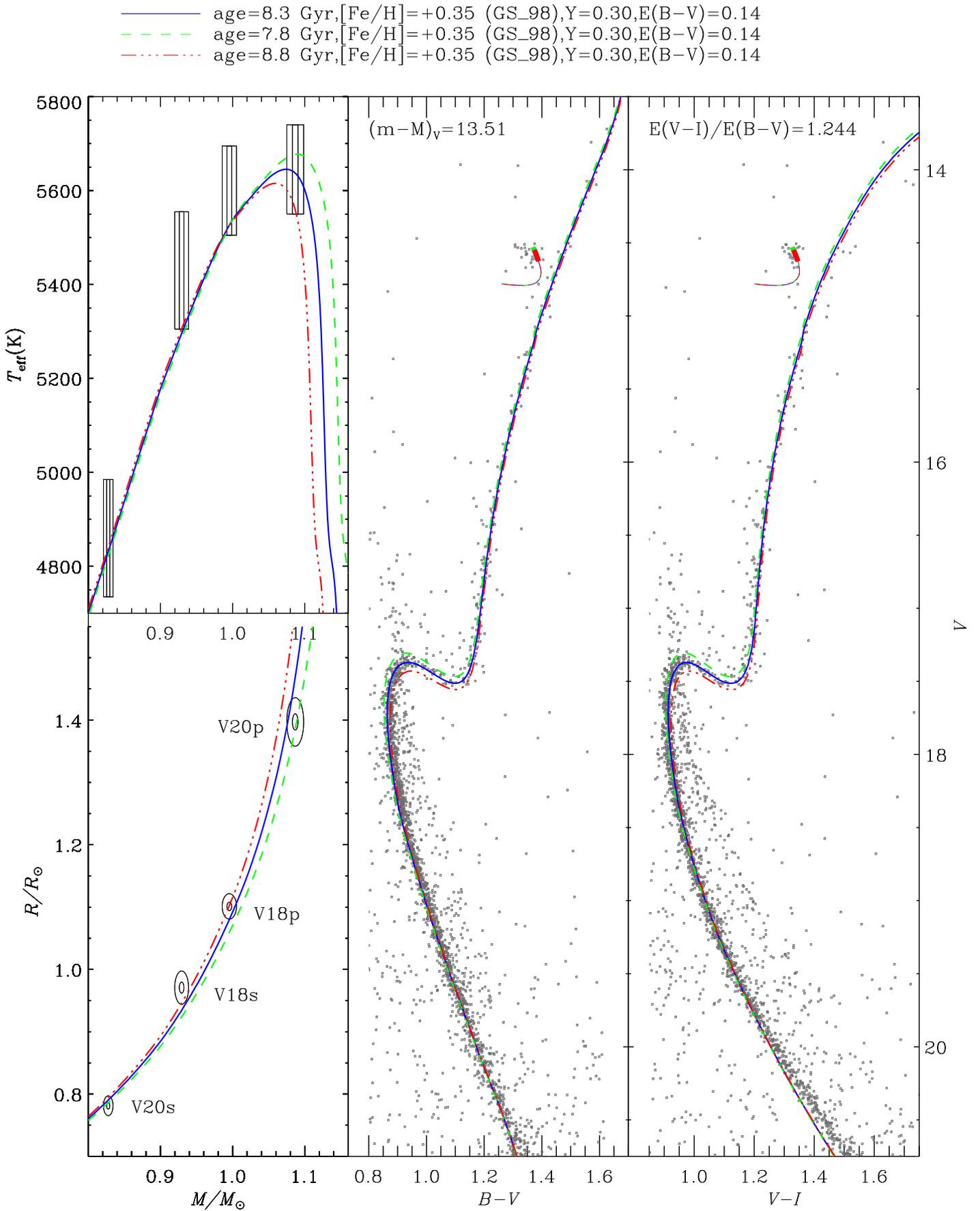}
\caption[]{\label{fig:measurement}
Measurements of the eclipsing binaries V18 and V20 (left panels) and cluster CMDs (middle and right panels) compared to Victoria model isochrones and ZAHB loci following the procedure described in Sect.~\ref{sec:proc}. Uncertainties shown in the left panels are $1\,\sigma$ and $3\,\sigma$ for mass and radius, and $1\,\sigma$ for $T_{\rm eff}$. GS\_98 denotes the solar abundance pattern of \cite{Grevesse98}. The thick part of the ZAHB corresponds to masses with an RGB mass-loss from 0 to twice the mean mass loss found from asteroseismology \citep{Miglio12}.
}
\end{figure*}

\begin{figure*}
\epsfxsize=190mm
\epsfbox{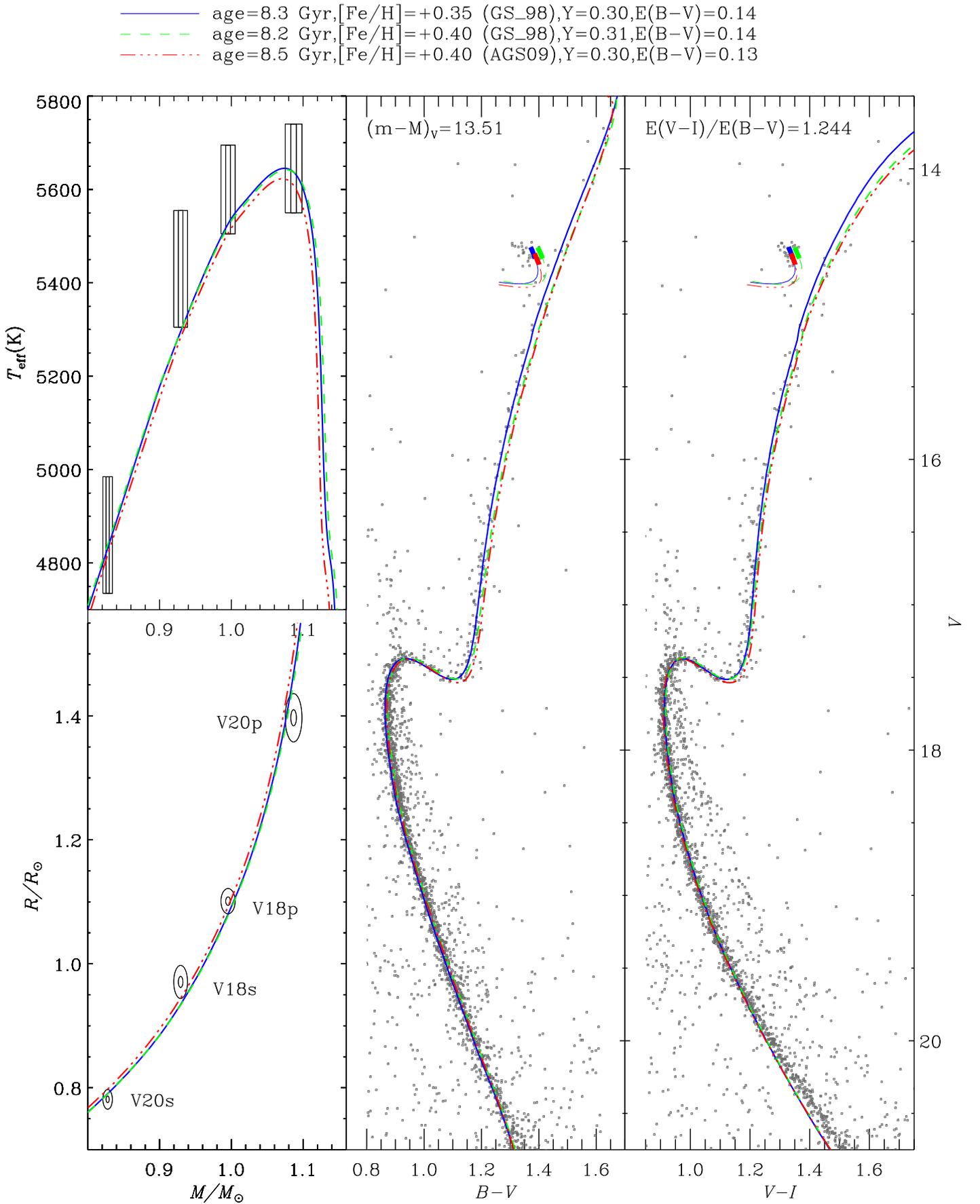}
\caption[]{\label{fig:degeneracy}
As Fig.~\ref{fig:measurement} but showing isochrones of different compositions. GS\_98 and AGS09 denote the solar abundance patterns of \cite{Grevesse98} and \cite{Asplund09}, respectively.
}
\end{figure*}

\begin{figure*}
\epsfxsize=190mm
\epsfbox{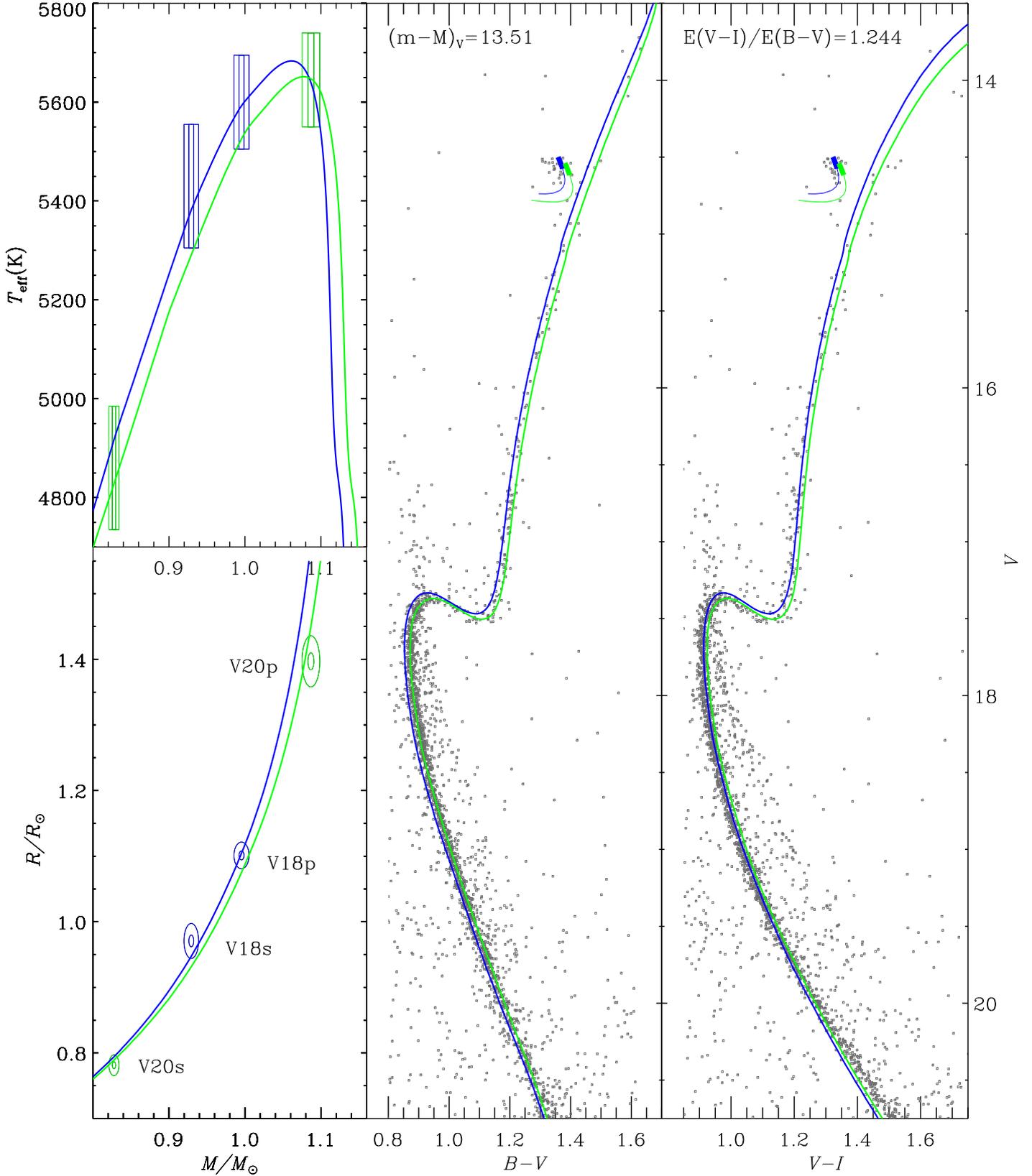}
\caption[]{\label{fig:deltafeh}
As Fig.~\ref{fig:measurement} but allowing for a small difference in [Fe/H] between the V18 and V20 systems, while the components of each system are still assumed to have identical [Fe/H]. Error boxes and ellipses are shown in the same colour as the isochrones meant to match them.}
\end{figure*}

\begin{figure*}
\epsfxsize=190mm
\epsfbox{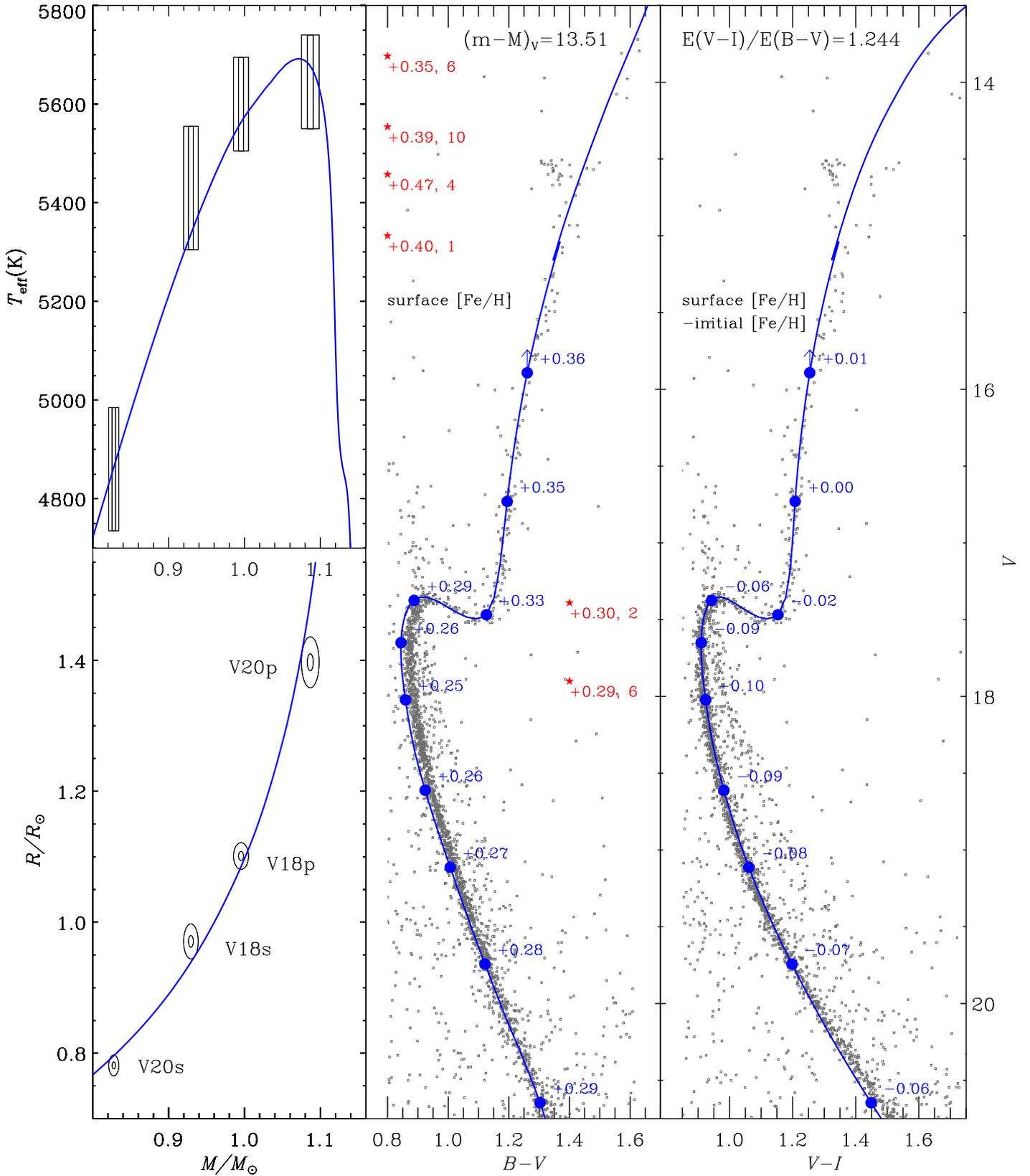}
\caption[]{\label{fig:metals}
As Fig.~\ref{fig:measurement} but showing a DSEP isochrone including heavy-element diffusion and turbulent mixing. In the middle panel the surface [Fe/H] values are shown at selected positions along the isochrone. Each star symbol marks the mean $V$-band magnitude of the stars in a spectroscopic investigation. The corresponding numbers are the measured [Fe/H] and the number of stars used. In order of increasing $V$-magnitude, the measurements are those of \cite{Origlia06}, \cite{Carraro06}, \cite{Gratton06}, \cite{Peterson98} plus the present paper, \cite{Boesgaard09}, and paper I.
In the right panel the change from initial to present surface [Fe/H] is shown at selected positions along the isochrone. } 
\end{figure*}

\begin{figure*}
\epsfxsize=195mm
\epsfbox{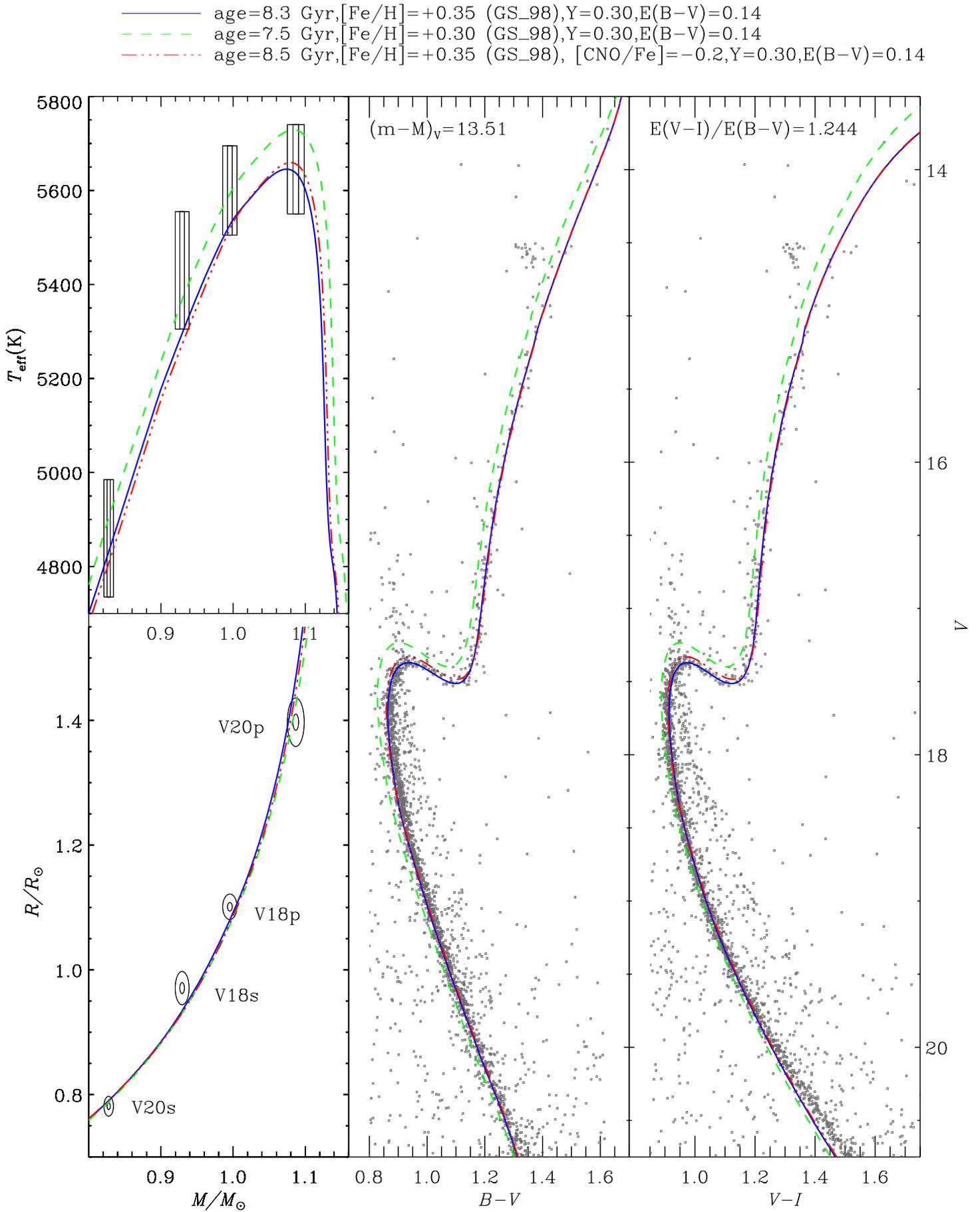}
\caption[]{\label{fig:cno}
As Fig.~\ref{fig:measurement} but showing isochrones of different compositions.}
\end{figure*}

Figs.~\ref{fig:measurement}-\ref{fig:cno} show our measurements of the binary components in mass-radius and mass-$T_{\rm eff}$ diagrams and the observed $V,B-V$ and $V,V-I$ CMDs compared to isochrones including ZAHB positions for stellar models with different compositions and differences in assumed physics. Here we will comment on the Figures, attempting to focus separately on the different factors at play and how the inferred age depends on them.

\begin{figure*}
\epsfxsize=195mm
\epsfbox{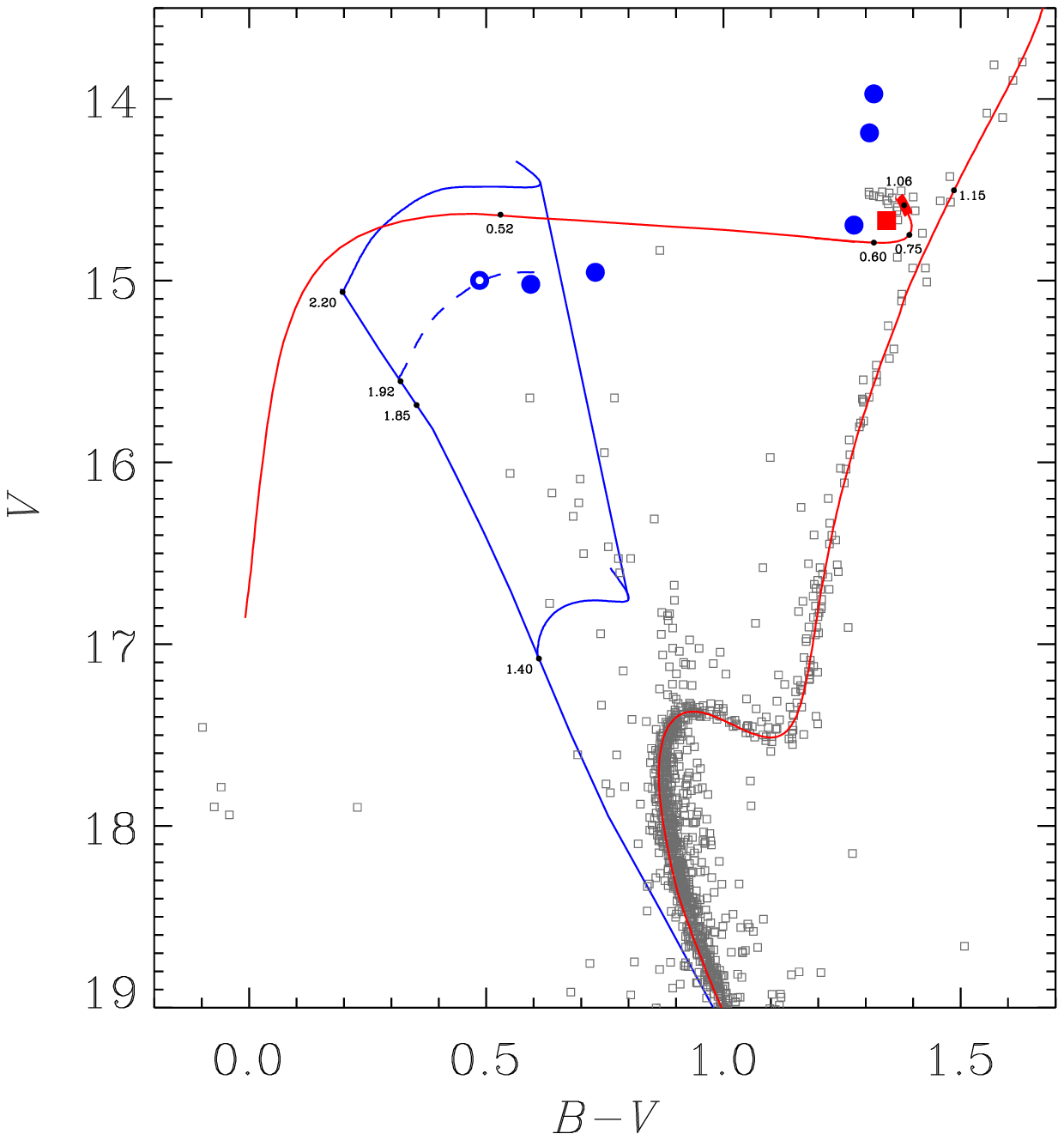}
\caption[]{\label{fig:bhbvsbss}
The $V,B-V$ CMD of NGC\,6791 compared to the isochrone and ZAHB from Fig.~\ref{fig:measurement} extended to bluer colours. Only stars with a non-zero proper motion probability in preliminary measurements of Kyle Cudworth (private comm.) are shown. Circles represent stars which were selected as suspected HB stars in the proper motion study by \cite{Platais11}. The open circle, which belongs to this group, is the star 2-17.
The square represents a cluster member that does not show the seismic signature of a RC star and has a seismic mass which is larger than that of the RGB stars \citep{Miglio12, Corsaro12}.
Also shown is an extended ZAMS and main-sequence evolutionary tracks for stars with masses of 1.4 and 2.2 $M_{\sun}$ taking into account convective core overshooting as in \cite{VandenBerg06}. The ends of the tracks are connected to show the terminal age main sequence (TAMS) in this mass interval.
The dashed line is the 2.2 $M_{\sun}$ track shifted down along the ZAMS to estimate the mass of 2-17.
Numbers indicate masses at the given positions.
}
\end{figure*}

\subsubsection{Measurement precision}
\label{sec:precision}

In Fig.~\ref{fig:measurement} we show a comparison of our binary measurements and CMDs to isochrones assuming $E(B-V) = 0.14$, abundances scaled from the solar abundance pattern of \cite{Grevesse98} to [Fe/H] = $+0.35$ and helium mass fraction $Y = 0.30$ while neglecting heavy-element diffusion. These assumptions allowed the overall best match to the observations. Isochrones are shown for the best matching age and $\pm 0.5$ Gyr. As illustrated, a very precise age of 8.3 Gyr is determined for this specific model with the adopted parameters. The acceptable age range seems to be very small when considering the mass-radius diagram as well as the SGB magnitude. However, the uncertainty of the binary $T_{\rm eff}$ measurements dominate the error in the measured distance modulus and these two measurements are therefore not independent. Because of this, one would expect the best model match to the $T_{\rm eff}$ measurements to also be the best match to the CMD using the measured distance modulus. In reality, we found that for a number of our models the best match to the SGB in the CMD only matched the mass-$T_{\rm eff}$ diagram at the lower $1\,\sigma$ limit of the effective temperatures. One could therefore instead adopt the approach of matching the mass-$T_{\rm eff}$ diagram as the second step of the comparison procedure, which corresponds to moving the $1\,\sigma$ disagreement from $T_{\rm eff}$ to the distance modulus. This would make the inferred age younger by up to 0.4 Gyr for some models, which limits the precision that can be reached for these models. At first sight one might think that the ZAHB position would help solve this ambiguity but it turns out that the slight composition change between the results from the two fitting approaches shifts the ZAHB luminosity such that it matches equally well in both cases. Since we do not adopt this latter procedure, we will shift downwards our final age estimate by 0.1-0.2 Gyr, depending on the consistency level of the given model, and adopt $\pm0.3$ Gyr as our measurement precision estimate.

\subsubsection{[Fe/H]}
\label{sec:feh}

A comparison of the solid and the dashed isochrones in Fig.~\ref{fig:degeneracy} shows that a correlation between [Fe/H] and the helium mass fraction $Y$ allows a range of acceptable model compositions which fit the observations equally well. With the present observations this ambiguity in [Fe/H] and $Y$ of the best fit is broken only if we trust that the models predict the colour of the red clump stars very accurately - something that is very questionable (see Sect.~\ref{sec:ct}). This means that the isochrones alone are not able to constrain [Fe/H], since we could produce isochrones with [Fe/H]$=+0.20$ or $+0.50$ that would match as well as the ones shown in Fig.~\ref{fig:degeneracy} by adjusting $Y$ along with [Fe/H]. It is therefore only because of our spectroscopic measurements in paper I and Sect.~\ref{sec:bhb} of this paper that we are able to constrain solutions to those with [Fe/H] in the range $+0.29 - +0.40$. 

Despite the ambiguity, we see from Fig.~\ref{fig:degeneracy} that the age is well constrained, since the effect on the age from changes in [Fe/H] is almost exactly compensated by the corresponding change in $Y$ needed to retain the match to the observations.

The age remains well constrained also when we assume abundances scaled from those of \cite{Asplund09} instead of those of \cite{Grevesse98}, as seen by comparing the dashed and dashed-dotted lines in Fig.~\ref{fig:degeneracy}. Any significant differences introduced by the change in the assumed solar abundance pattern are compensated by the solar calibration and by a slight difference in helium content. Our observations therefore do not give preference to any particular set of solar abundances. On the positive side this means that the inferred age does not depend upon the particular choice of solar abundances.

The comparisons so far assumed that the two eclipsing binaries have identical initial chemical composition. In Fig.~\ref{fig:deltafeh} we show that if we allow for a difference of 0.05 dex in [Fe/H] between V18 and V20 then we can obtain a much better match in the mass-radius and mass-$T_{\rm eff}$ diagrams. A corresponding [Fe/H] variation in the CMDs is seen to be at the upper limit of what can be allowed by the main sequence colourspread, while the magnitudespread of the SGB remains in accordance with observations. The observed RGB is wide enough to accommodate the predicted colourspread, and the observed colour variation among the RC stars is also better explained in this scenario. The presence of a star-to-star variation in [Fe/H] of 0.05 dex or less could therefore be real. This is, in fact, the only scenario that can match the measurements in the mass-radius diagram to the statistically expected level, assuming that the measurement uncertainty estimates are correct. However, since this requires an additional free parameter for which there is presently no conclusive evidence, and since there are many examples in the literature where mass and radius measurements changed at the $1-2\,\sigma$ level as better observations became available, we consider this scenario only as a possibility until further evidence becomes available. The [Fe/H] measurements of the individual binary components are not precise enough to be used for such evidence.
If accompanied by a correlated variation in $Y$, then [Fe/H] variations could even be larger as was seen in Fig.~\ref{fig:degeneracy}. Nevertheless, as seen in Fig.~\ref{fig:deltafeh}, a star-to-star variation in [Fe/H] has no significant effect on the inferred cluster age.

\subsubsection{Heavy element diffusion}
\label{sec:metals}

In Fig.~\ref{fig:metals} we show an isochrone corresponding to a DSEP model including heavy element diffusion and turbulent mixing with a reference temperature $\log\,T = 6.0$. As seen, the inclusion of heavy element diffusion in the models does not affect the inferred age significantly (compare to Fig.~\ref{fig:degeneracy}). 

In the $B-V$ plane in Fig.~\ref{fig:metals} we show the surface [Fe/H] values at selected positions along the isochrone with initial [Fe/H] = $+0.35$. The effect of heavy element diffusion is seen to be a depletion of the surface [Fe/H] on the main sequence, reaching about 0.1 dex lower than the initial abundance at the turn-off. When the stars leave the main sequence, the convection zone deepens and brings the metals back up into the atmosphere, causing the surface [Fe/H] to be very similar to the initial value on the RGB. The surface [Fe/H] becomes slightly larger than the initial value on the RGB because the convection zone reaches depths where hydrogen has been partly processed into helium.
The model with $\log\,T = 6.0$ is arguably the currently best description of heavy element diffusion, since, with this description of turbulence, a model can reproduce abundance measurements in the globular cluster NGC\,6397 \citep{Korn07} as well as other constraints \citep{Richard05}. However, since the turn-off stars in NGC\,6791 are cooler and have thicker convective envelopes than their counterparts in the globular clusters, the turbulent mixing is much less important here. Indeed, we found that removing the turbulent mixing completely from the models does not significantly change the isochrone or predicted surface [Fe/H] values. At the other extreme, when we adopt a very high value for the turbulence reference temperature $\log\,T = 6.5$, the turn-off surface [Fe/H] is still depleted by 0.06 dex, in good agreement with the conclusion of \cite{Proffitt91} for the maximum effect of turbulent mixing on diffusion. On top of this, we found that the surface [Fe/H] depletion is not changing significantly with the assumed initial [Fe/H]. Therefore, we show on the $V-I$ plane of Fig.~\ref{fig:metals} a rather robust prediction for the [Fe/H] surface abundance change along the cluster sequence, which can be compared to observations.

\cite{Twarog10} noted that the the lowest measured [Fe/H] value in the literature (at that time) was that found for early SGB stars by \cite{Boesgaard09} while other high-resolution [Fe/H] measurements from stars in later evolutionary stages all give higher values, as shown in Fig.~\ref{fig:metals}. This suggests that the signature of heavy element diffusion might already have been observed. While large measurement uncertainties might also play a role, it is interesting that this pattern remains when we add our [Fe/H] measurement of turn-off and early SGB stars from paper I and that from a high-mass blue straggler in Sect.~\ref{sec:bhb}. However, more homogeneous and higher quality measurements are needed to show whether these indications are real.

Our model comparisons give us a mixed message when it comes to the reality of heavy-element diffusion. The effective temperatures on the main-sequence are higher when including heavy-element diffusion, partly due to a larger mixing length from the solar calibration and partly due to a lower surface opacity. This improves the match in the mass-$T_{\rm eff}$ diagram. But in the CMDs, the combined effect of higher temperatures and lower surface [Fe/H] shifts the isochrones to the blue. This effect is particularly strong in $B-V$, because this index is very sensitive to the surface [Fe/H] value.  
Since we have already chosen $E(V-I)/E(B-V)$ at the extreme lower limit, we cannot choose a higher reddening $E(B-V)$ in order for the isochrone to match the $V,B-V$ CMD while keeping a good match to the $V,V-I$ CMD. If we were able to trust the colour-temperature transformations, this would indicate that heavy-element diffusion in real stars is not proceeding the way it is described in current stellar models. Unfortunately we cannot trust colour-temperature transformations to this level and therefore the much weaker conclusion is that either we do not properly understand heavy-element diffusion or the colour-temperature transformations are deficient, with the latter explanation being the most likely one. 

It is interesting to note that since models including heavy-element diffusion require a higher value of the mixing length parameter to match the Sun, the isochrones also become hotter and therefore bluer on the giant branch. Because of this, the dominant effect is an overall shift in colour while the shape of the isochrones in the CMDs change very little, except at cooler effective temperatures, where uncertainties in e.g. surface boundary conditions (\citealt{VandenBerg08}), TiO molecular lines (\citealt{Garnavich94,Carney05}), and helium content (\citealt{Girardi07}) makes the interpretation difficult. It is therefore very problematic to use only the CMD to determine cluster parameters from an overall best fit isochrone.
 
\subsubsection{CNO abundances}
\label{sec:cno}

The initial CNO abundances in NGC\,6791 are presently not well known, with large differences
 among published measurements (\citealt{Origlia06, Carretta07, Peterson98}), ranging
from [CNO/Fe] $\sim-0.3$ to solar [CO/Fe] and very high [N/Fe]. In addition, these measurements were performed on evolved stars and some of the authors find indications of extra-mixing processes on the RGB making it difficult to relate the measured values to the initial abundances.  

To investigate the effect that the assumed initial [CNO/Fe] has on the age of stars in NGC\,6791, we computed an isochrone assuming $\mathrm{[CNO/Fe]}=-0.2$, which we present in Fig.~\ref{fig:cno} together with isochrones assuming a scaled solar composition. Comparison of the solid and dashed lines shows that, for a scaled solar composition, the age decreases, temperatures increase and the SGB brightens when [Fe/H] is decreased at a fixed mass-radius relation. Comparing instead the solid and dashed-dotted lines shows a very different picture when only CNO abundances are decreased; the age increases while the SGB brightens without a significant temperature change on the main-sequence. Since a slight increase in [Fe/H] or decrease in $Y$ is needed in order to shift the SGB magnitude in agreement with observations for the CNO reduced isochrone, the effect on age is an increase of about $0.5$ Gyr. The uncertainties in predicted $T_{\rm eff}$s caused by the possible effects of heavy-element diffusion and star-to-star variations in [Fe/H] limits the use of the measured $T_{\rm eff}$ for conclusions about the reality of CNO underabundances as well as useful constraints to their values. 

\subsubsection{The mixing length parameter}
\label{sec:mr}

The mixing length parameter of our models is held fixed at the value found for solar calibrations, and we have thus not explored the effects of allowing the mixing length to vary with stellar parameters or evolutionary stage. In the following we examine the potential consequences of this limitation. 
\cite{Trampedach11} made theoretical predictions for the variation of the mixing length as a function of $T_{\rm eff}$ and $\log g$ based on 3D convection simulations. To see how such predictions would affect our results, we make use of their Fig. 6 to obtain a value of the mixing length for each of our four binary components and the Sun. Since the solar calibrated mixing length for our models is higher than the solar value predicted by \cite{Trampedach11} we do not use their absolute numbers, but calculate instead their predicted offset from the solar value for each star. Since the two most massive, and therefore most age-dependent binary components happen to have almost exactly the same predicted value of the mixing length as the Sun, employing such offsets would not affect the inferred age and helium content for a given model comparison. To test the effect of a varying mixing length, it is therefore sufficient to calculate the predicted radii of the two least massive binary components at the age of one of our best matching models using mixing lengths obtained by adding the predicted offsets to the solar calibrated value. The results are shown in Table~\ref{tab:alpha}. As can be seen, the effect is to lower the predicted radii and to increase the predicted $T_{\rm eff}$, but not by amounts which significantly affects the quality of the match to the observations since changes are at the same level or smaller than the measurement uncertainties. 

\begin{table}   
\begin{center}
\caption[]{\label{tab:alpha}
Effect of varying the mixing length parameter with effective temperature and surface gravity.
}
\begin{tabular}{lcccc} \hline    
\hline
\noalign{\smallskip}    
			& $M(M_{\sun})$	& $\alpha _{MLT}$	& $T_{eff}(K)$	&	$R(R_{\sun})$\\
\hline    
\noalign{\smallskip}    
V20s			&		&		&		&	\\
Measurement		& 0.8276	& -		& 4860		&	0.7813\\		
Measurement uncertainty(1$\sigma$)	& 0.0022& -		& 125		&	0.0053\\
\noalign{\smallskip}    
Model - fixed $\alpha _{MLT}$	& 0.8276	& 2.005		& 4837		&	0.789\\		
Model - variable $\alpha _{MLT}$& 0.8276	& 2.265		& 4889		&	0.779\\
change			& -		& +0.26		& +52		&	-0.010\\		
\noalign{\smallskip}    
\hline
V18s			&		&		&		&	\\
Measurement		& 0.9293	& -		& 5430		&	0.9708\\		
Measurement uncertainty(1$\sigma$)	& 0.0032& -		& 125		&	0.0089\\
\noalign{\smallskip}    
Model - fixed $\alpha _{MLT}$	& 0.9293	& 2.005		& 5302		&	0.933\\		
Model - variable $\alpha _{MLT}$& 0.9293	& 2.075		& 5323		&	0.927\\
change			& -		& +0.07		& +21		&	-0.006\\		

\noalign{\smallskip}    
\hline    
\end{tabular}            
\end{center}
\tablefoot{For the model with fixed mixing length the value was obtained from a solar calibration. The models with variable mixing length has added to this value a difference predicted using Fig. 6 of \cite{Trampedach11}.} 
\end{table}

The $T_{\rm eff}$ and thus colour of the RGB is very sensitive to the precise value of the mixing length. However, looking again at Fig. 6 of \cite{Trampedach11}, we see that the predicted difference in mixing length between the most massive binary component and later evolutionary stages is of the order of only -0.02 or less. While adopting the predicted variation in mixing length would shift the predicted colour of the RGB, the magnitude of the change is much smaller than the uncertainty of the observed RGB position. A variable mixing length would make the slopes of isochrones steeper on the main sequence in the CMDs, but the large uncertainties in the colour-temperature transformations makes that a weak indicator as well.

In conclusion, the relatively small consequences of adopting the variations in mixing length predicted by \cite{Trampedach11} leaves us unable to distinguish between the case of a fixed or a varying mixing length. In comparison to our measurement precision, the isochrones are simply not significantly affected by a variable mixing length, unless variations are larger than, or different from, the predictions of \cite{Trampedach11}. Measurements of more cluster member eclipsing binaries with lower masses could improve this situation in the future.

\subsubsection{Colour-temperature transformations and the reddening}
\label{sec:ct}

Our inability to accurately predict the relation between colours and effective temperatures of stars is a major uncertainty when comparing observations to stellar models in the CMD --- in particular at super-solar metallicities. In our model comparisons we have chosen colour-temperature transformations derived from MARCS 1D atmosphere models. However, there are significant differences between different up-to-date colour-temperature relations, as demonstrated by \cite{VandenBerg10}. Therefore, the overall fairly good match between models and the observed CMD is likely to be more of a coincidence than a reflection of reality and we caution the reader that one should not trust the CMD colours to identify the most realistic stellar model.

As an example, \cite{VandenBerg10} noted that changing the value of the microturbulent velocity $\epsilon$ in the MARCS atmosphere models from 1 km/s to 2 km/s results in a change of ~0.02 mag. in the $B-V$ colour of a star with solar parameters, while colours at longer wavelengths, not including the $B$ filter, are not significantly affected. The direction of this shift, which was not reported, is to redder colours, e.g. larger values of $B-V$.  
In paper I we found from our spectroscopic analysis that $\epsilon \sim 1$ km/s for the turn-off stars in NGC\,6791. The MARCS based colour-temperature relations we use, assume $\epsilon =2$ km/s. If we demand consistency with the spectroscopic result, we would need to shift the predicted $B-V$ colours by $\sim0.02$ mag to the blue, assuming the offset found by \cite{VandenBerg10} is roughly independent of $T_{\rm eff}$ and $\log g$. Since Figs.~\ref{fig:measurement} -~\ref{fig:cno} already assume an extreme lower limit for the ratio $E(V-I)/E(B-V)=1.244$, it is obvious from these figures that such a shift would not allow a consistent match to the CMDs in both colour planes, since adjusting $E(B-V)$ to retain agreement with observations in $B-V$ would make the predicted $V-I$ colours too red. The apparent inconsistency in the choice of microturbulent velocity exemplifies the magnitude of uncertainty present in stellar atmosphere models; Table 7 in \cite{Casagrande10} reveals that the predicted $B-V$ colour of the Sun is redder by about 0.02 for ATLAS9 model atmospheres compared to MARCS at the same assumed microturbulent velocity. One might therefore suspect that using colour-temperature relations based on ATLAS9 model atmospheres would remove the inconsistency. Even so, the colours remain problematic, particularly for the models including heavy-element diffusion.

The adoption of empirical colour-temperature relations does not lead to a resolution of the problem. We found, interestingly, that the empirical $B-V$ relations of \cite{Sekiguchi00}, \cite{Ramirez05}, and \cite{Twarog09} predict almost exactly the same shape of the SGB, the turn-off region and the upper main-sequence on the $B-V$ plane, but with colour offsets at the turn-off of $-0.015$, 0, and +0.025 mag, respectively, relative to MARCS transformations. The predicted main sequence slope is shallower than observed, becoming worse for the isochrones including heavy-element diffusion. The $B-V$ transformations of \cite{Casagrande10} match those of MARCS at the SGB and the turn-off, but predicts a much shallower main sequence than any of the other transformations, leading to the worst match to the $B-V$ data. At the same time, however, the transformations of \cite{Casagrande10} lead to an excellent match of the shape of the whole main sequence, turn-off and SGB on the $V-I$ plane, in particular for isochrones not including heavy-element diffusion.

Because of the big uncertainty related to the $B-V$ colours, our best estimate of the reddening, $E(B-V)$, relies on $V-I$ rather than $B-V$. This makes the derived value of $E(B-V)$ dependent on the ratio $E(V-I)/E(B-V)$ which is not known precisely. For reasonable values around 1.35 (\citealt{Hendricks12}, Tables 11-13) we find $E(B-V)=0.12-0.15$ from our isochrone fits to $V-I$ which can be made consistent also with $B-V$ if we allow a systematic shift of $+0.02$ or more in the predicted $B-V$ colour. We adopt $E(B-V)=0.14\pm0.02$ as our best reddening estimate.

The uncertainties in the colour-temperature transformations renders the CMDs problematic for a precise age measurement because they make it impossible to use the quality of the model fit to the CMD to select among different model assumptions. Luckily we can come a long way by just relying on the $V$-band magnitude of the SGB, which is the only feature of the CMDs which is used in our age estimates. This still relies on a bolometric correction (BC) in the $V$-band, which we get from MARCS atmosphere models. However, since the same source is used for the BC of the binary components when calculating the distance modulus, we are only concerned about the relative error in the BC over a small interval in $T_{\rm eff}$ and $\log g$, which should be very small.

\subsubsection{Binarity effects}
\label{sec:sage}

The assumption of the model comparisons we are making is that the binary star components evolve as single stars and are not significantly affected by the binarity. For binaries of lower mass and very short orbital periods this is often not true, with binary components being larger and cooler than predicted by single-star evolution, currently interpreted as the efficiency of convection being inhibited by strong magnetic fields induced by fast rotation.
Although our binary stars have relatively long orbital periods (P$\sim$15 and 20 days) one could still be concerned whether they are affected similarly, although to a lower extent. While it is presently impossible to rule this out, there are a number of indications which suggests that such effects are not significant; In our model comparisons the measured temperatures of the binary components are always equal to or larger than predicted, while the effects of binarity should make them cooler. The stellar radii are not correlated with rotation if one assumes that rotation is close being to synchronized with the orbit; V18 orbits slower than V20 and yet is the system where the radii of the components are slightly larger than predicted, while the opposite is true for V20. Binarity effects should be larger at lower mass \citep{Clausen09} but the isochrones match well both the components of V20, which have the lowest and highest masses measured. 
Finally, the two low-mass binary components are always matched quite well despite that our best matching isochrones were found using only the two most massive components. This suggests that binarity effects are small, if present at all. We thus believe that not taking into account binarity has a negligible effect on our age predictions in comparison to other uncertainties.

\subsubsection{A prolonged star-formation?}
\label{sec:pro}

\cite{Twarog11} put forward an idea that the increased scatter in the turn-off region of CMDs of NGC\,6791 could be caused by a prolonged star-formation lasting up to 1 Gyr. Their argument is based on the finding that CMDs of the inner and outer parts of the cluster show turn-off regions which are offset in colour relative to each other. The observed SGB is, however, much too thin to accommodate an age spread of 1 Gyr for a single metallicity case, see Fig.~\ref{fig:measurement}. The colour offset is also easily explained by differential reddening and goes away under such an assumption, cf. Sect.~\ref{sec:difred}. 
Experiments with isochrones as those in Fig.~\ref{fig:deltafeh} nevertheless show that under the assumption of small star-to-star variations in [Fe/H] it is possible to find solutions where the most metal-poor stars are up to 0.3 Gyr older than the most metal-rich while keeping the SGB thin in the CMDs. If accompanied by a variation from [CNO/Fe]= $-0.2$ for the most metal-poor stars to [CNO/Fe]= 0.0 for the most metal-rich then an age variation of as much as 0.7 Gyr would be possible without being revealed in the CMDs, as can be inferred by combining results from Figs.~\ref{fig:deltafeh} and ~\ref{fig:cno}. Although this latter scenario requires fine-tuning, we emphasize the importance of determining whether or not intrinsic abundance variations are present among the cluster stars before a prolonged star-formation can be ruled out.

\subsubsection{Helium abundance and inferences for $\Delta Y/\Delta Z$}
\label{sec:helium}

As seen in Figs.~\ref{fig:measurement}-\ref{fig:cno} the helium mass fraction $Y$ depends only weakly on the exact [Fe/H], the assumed solar abundance pattern, and [CNO/Fe], and serves in most cases to compensate the effect that the other assumptions have on age. For all cases within our spectroscopic [Fe/H] constraints we have $Y=0.30\pm0.01$. The uncertainty in $Y$ due to our measurement procedure, as described in Sec.~\ref{sec:proc} and \ref{sec:precision}, is at the level of $\pm0.005$ or less.

Assuming that helium enrichment follows the enrichment of heavy elements in a universal way and can be parametrized as $\Delta Y/\Delta Z$, we can use our model comparisons to put constraints on this helium-enrichment parameter.
We calculated $\Delta Y/\Delta Z$ for a subset of our best matching models both from initial solar values to initial values in NGC\,6791, from primordial to initial solar values, and from primordial to initial values in NGC\,6791. For the primordial value we used $Y=0.248$ (e.g. \citealt{Peimbert07, Cyburt08}).

Interestingly, when including heavy element diffusion, the three calculations give almost the same value, suggesting a linear $\Delta Y/\Delta Z$ all the way from primordial to the high metallicity of NGC\,6791. The exact value of $\Delta Y/\Delta Z$ depends on the exact [Fe/H] and to a lesser extent on the assumed abundance pattern, but the values range from $\sim 1.3$ to $1.5$ when heavy element diffusion is accounted for, so our present best estimate for $\Delta Y/\Delta Z$ is $1.4\pm0.1$.
Models neglecting heavy element diffusion do not reproduce this result, giving a steeper slope from solar metallicity and up, rising from 1.0 (primordial to solar) to 2.0 (solar to NGC\,6791). 

\subsubsection{Best age estimate}
\label{sec:bage}

Taking the results of this section as a whole, we found that the age of NGC\,6791 is very well constrained for given assumptions of physical ingredients and composition.
In addition, the predicted age is insensitive to the exact value of [Fe/H] and whether one relies on the solar abundance pattern of \cite{Asplund09} or \cite{Grevesse98}, as long as a scaled solar composition is assumed for NGC\,6791.
Effects on age from possible binarity effects, variable mixing length, colour-temperature relation errors, and star-to-star variations in [Fe/H] and age have been shown or argued to be small or negligible. However, the combination of uncertainties related to these issues makes it impossible to make firm conclusions about the presence of heavy-element diffusion and/or non-solar [CNO/Fe] in NGC\,6791 without further observational evidence. 

Our present best age estimate is that of a model including heavy element diffusion, giving more weight to the simultaneous match of the effective temperatures and distance moduli of the binaries and the ability to produce a linear helium enrichment law than questionable indications from colours in the CMD. Due to the uncertain and conflicting results of present CNO measurements as well as their uncertain relation to the initial CNO abundances we assume for now scaled solar abundances. 

After applying an offset based on Sect.~\ref{sec:precision} our present best estimate for the age of NGC\,6791 is thus $8.3\pm0.3$ Gyr, where the quoted uncertainty is that originating from the fitting process. To this uncertainty we have to add additional contributions, where the questionable reality of heavy element diffusion adds about $\pm0.2$ Gyr, while a potential prolonged star-formation could shift the age estimate by about $+0.2$ Gyr. The uncertainty in the CNO abundances is now the major factor in the final error budget. Using a rough scaling of our results, our age estimate will increase by approximately 0.7 Gyr if [CNO/Fe]$\sim-0.3$ \citep{Carretta07} are representative of the initial values for NGC\,6791. 

\subsection{New cluster insights}
\label{sec:new}

\subsubsection{The horizontal branch, RGB mass and mass-loss}
\label{sec:massloss}

In Fig.~\ref{fig:bhbvsbss} we show the CMD of NGC\,6791 compared to one of our best matching isochrones and corresponding ZAHB {with masses indicated at selected positions}. As seen, the horizontal branch stars in the RC of NGC\,6791 actually occupy a part of the horizontal branch which is not horizontal in CMDs with $V$-magnitude on the Y-axis. The magnitudes and the presence of a sloped lower envelope of the RC stars (even more evident in $V-I$ CMDs in Figs.~\ref{fig:measurement}-\ref{fig:cno}) is evidence that they follow the upturn predicted by the theoretical ZAHB, except for a colour-shift.

Tests with MESA \citep{Paxton11} horizontal branch evolutionary tracks for masses in the range $0.8-1.2 M_{\sun}$ predict that stars remain close to their ZAHB position for most of their HB lifetime, and then evolve quickly towards higher luminosity and lower $T_{\rm eff}$ corresponding to redder colours. Therefore, RC stars with masses above $\sim0.75 M_{\sun}$, the beginning of the ZAHB upturn, should be on the ZAHB upturn or slightly redder. This suggests that our models predict too red colours, since the RC stars are on the blue side of the predicted ZAHB upturn. Interestingly, the $T_{\rm eff}$ and therefore colour at the upturn of the ZAHB, is very sensitive to the stellar abundances, in particular that of Oxygen, providing a potential indirect way of constraining the Oxygen abundance, if other colour uncertainties can be overcome.

Our model ZAHB matches very well the mass-range in the RC which is inferred using the RGB mass-loss of $\Delta M \sim 0.09 M_{\sun}$ from asteroseismology \citep{Miglio12}. The fact that the absolute masses that we get on the RGB and in the RC are smaller than the corresponding values from seismology \citep{Basu11,Miglio12} is not so surprising, given the rather big uncertainties, and errors of systematic character, which could be present in the seismic measurements at this point. 

All our acceptable model fits give a mass on the RGB at the $V$-magnitude of the RC in the range $1.15\pm0.02 M_{\sun}$. Our isochrones do not include mass-loss, so in principle the RGB mass could be lower, but it is unlikely that a significant mass-loss has occurred at the relatively low luminosities below the RC. 

\subsubsection{Reclassification of suspected blue horizontal branch stars}
\label{sec:bhb}

Stars marked with circles in Fig.~\ref{fig:bhbvsbss} were selected as suspected horizontal branch stars in the proper motion study by Platais et al. (2011). As is evident, the hottest of these stars are significantly fainter than predicted for ZAHB stars. [Fe/H] values are correlated with the helium content in our best estimate models, and changing these two parameters shifts the ZAHB in opposite direction in luminosity. Therefore, we cannot find any acceptable model for which the ZAHB would match these stars. If we ignore this and apply an ad-hoc magnitude offset to match the blue stars, the ZAHB would become too faint to match the stars in the RC. Thus, the stars must be blue straggler stars (BSS) rather than blue horizontal branch (BHB) stars, unless there is something seriously wrong with our theoretical understanding of the horizontal branch.

To show that the BSS scenario is allowed by our model predictions, we show in Fig.~\ref{fig:bhbvsbss} an extended ZAMS and main-sequence evolutionary tracks for models with masses of 1.4 and $2.2 M_{\sun}$, the latter corresponding to twice the turn-off mass, which is well-defined due to our eclipsing binary measurements. Since the $2.2 M_{\sun}$ track is more luminous than the stars in question, they could have been produced by the merger of two normal main-sequence stars.

For one of the stars, known as 2-17 \citep{Kinman65}, marked with an open circle in Fig.~\ref{fig:bhbvsbss}, \cite{Peterson98} carried out a spectroscopic analysis. They concluded that the star could be either a BSS or a BHB star, but for unknown reason the latter option was the only one which was noted in most later papers which cited their result (e.g. \citealt{Origlia06, Carretta07, Carraro06, Carney05, Gratton06, Boesgaard09}; mentioning just a few).
We performed a new spectroscopic analysis of the spectrum obtained by \cite{Peterson98}. For details on the spectrum we refer to \cite{Peterson98}. The analysis method was the same as for the binary stars in paper I, except that we did not use any prior information on $\log g$. We used 46 \ion{Fe}{I} and 7 \ion{Fe}{II} lines with equivalent widths between 25 and 100 milli-Angstrom in the wavelength range 5300-6600 Angstrom. Since this is a hot star we employed NLTE corrections from \cite{Rentzsch-Holm96}. However, our results and conclusions remain unchanged if we neglect the NLTE corrections, since in that case the inferred $\log g$ and [Fe/H] are both reduced in each solution by 0.12 dex and 0.07 dex respectively.

The analysis shown in Table~\ref{tab:217} revealed that due to the correlation between $T_{\rm eff}$ and $\log g$, a large range of acceptable solutions exist when no additional information is used. However, by combining the spectroscopic $T_{\rm eff}$ with the the $V$-band magnitude of the star and the precise distance modulus from our binary star measurements, we can calculate the radius of the star, and therefore either the mass needed to obtain agreement with the spectroscopic $\log g$, or an alternative $\log g$, in each case. As seen in Table~\ref{tab:217}, the mass turns out to be higher than the turn-off mass in all cases, strongly indicating that this star is a blue straggler. 
Assuming that the evolution of a BSS can be approximated by that of a normal star, we obtain from the models in Fig.~\ref{fig:bhbvsbss} a mass of about $1.92 M_{\sun}$, whereas a mass $\lesssim0.60 M_{\sun}$ is required for the BHB scenario.
Table~\ref{tab:217} shows that, for the latter option there is no solution with a consistent value of $\log g$.
In order to reach consistency in even the most favorable case, an unrealistically large error in the distance modulus or the bolometric correction of $\sim0.9$ mag, more than 10 times their estimated uncertainty, would be needed.
The discrepancy is large for all BHB solutions and is largest for the solutions with high [Fe/H]. This strongly suggests that the star is not on the BHB. Furthermore, it makes it impossible to match a ZAHB to the star 2-17 by invoking large star-to-star variations in [Fe/H] and a mass-loss which increases strongly with [Fe/H], as suggested by \cite{Kalirai07}.

Assuming instead a mass of $\sim$\,1.92 $M_{\sun}$ allows a solution with consistent values of $\log g$ while also being in good agreement with the other measurements of the cluster, [Fe/H] and photometric $T_{\rm eff}$. The MARCS $B-V$ colour-transformation predicts too blue colours, but this is in accordance with the problems found for the isochrones in this colour plane. The best solution is seen to be somewhere between solutions 3 and 4 in Table~\ref{tab:217}, resulting in [Fe/H]= $+0.40\pm0.08$, the same value as found by \cite{Peterson98}. At the lower uncertainty-limit of this measurement, a significant slope of FeI with excitation potential appears, which is the reason we do not consider stellar models with [Fe/H] below $+0.3$ in any of our model comparisons.

\cite{Peterson98} measured relative abundances [C/Fe], [N/Fe] and [O/Fe] at or above zero for 2-17, while these quantities have been measured to be sub-solar for giant stars in NGC\,6791 (\citealt{Carretta07, Origlia06}). If this difference is real, and not caused by large measurement errors, then it is another indication that star 2-17 did not follow a standard single-star evolution, further supporting that star 2-17 is a BSS and not a BHB star.

Considering the results of this section, star 2-17 and the other blue stars at the same luminosity level, suggested by \cite{Platais11} to be BHB stars, are almost certainly BSSs. In fact, since the proper motion study by \cite{Platais11} is expected to include all bright cluster members and yet contains only one other bright blue star, except for those on the EHB, it is very unlikely that there are any regular BHB stars present in this cluster. 

\begin{table*}   
\begin{center}
\caption[]{\label{tab:217}
Analysis of the hot star 2-17.}
\begin{tabular}{cccccc|cccccc} \hline    
\hline
\noalign{\smallskip}    

N & \multicolumn{2}{c}{[Fe/H] from} & $T_{\rm eff}(K)$ & $\log g$ & Excitation & $\Delta$ [Fe/H]\tablefootmark{a}& $\Delta (B-V)_0$\tablefootmark{b} & $\Delta (V-I)_0$\tablefootmark{b} & \multicolumn{2}{c}{$\Delta \log g$\tablefootmark{c} assuming 2-17 belongs to} & $M(M_{\sun})$\tablefootmark{f}\\

 & FeI & FeII &  &  & slope? & &  &  & BSS$(1.92M_{\sun})$\tablefootmark{d} & BHB$(0.6M_{\sun})$\tablefootmark{e} & \\

\hline    
\noalign{\smallskip}    
1 & $+0.54$ & $+0.51$ & 7400 & 4.18 & \bf{NO} 		& $+0.25$ & $-0.05$ & $-0.06$ & $+0.29$ & $+0.80$ & 3.75\\
2 & $+0.47$ & $+0.44$ & 7300 & 4.03 & \bf{NO} 		& $+0.18$ & $-0.04$ & $-0.04$ & $+0.17$ & $+0.67$ & 2.81\\
\emph{3} & $\mathit{+0.48}$ & $\mathit{+0.47}$ & \emph{7200} & \emph{3.91} & \bf{\emph{NO}} 		& $\mathit{+0.19}$ & $\mathit{-0.02}$ & $\mathit{-0.02}$ & \bf{\emph{$+$0.07}} & $\mathit{+0.57}$ & \emph{2.25}\\
\emph{4} & $\mathit{+0.32}$ & $\mathit{+0.30}$ & \emph{7100} & \emph{3.68} & $\mathit{+1.1 \sigma}$ 	& \bf{\emph{$+$0.03}} & $\mathit{-0.02}$ & \bf{\emph{$+$0.00}} & \bf{\emph{$-$0.13}} & $\mathit{+0.37}$ & \emph{1.42}\\
5 & $+0.26$ & $+0.27$ & 7000 & 3.62 & $+1.6 \sigma$ 	& $\mathbf{-0.03}$ & $-0.01$ & $+0.02$ & $-0.16$ & $+0.34$ & 1.31\\
6 & $+0.18$ & $+0.18$ & 6920 & 3.58 & $+2.5 \sigma$ 	& $-0.11$ & $\mathbf{+0.00}$ & $+0.04$ & $-0.18$ & $+0.33$ & 1.27\\
\noalign{\smallskip}    
\hline    
\end{tabular}            
\tablefoot{Table shows spectroscopic solutions N = 1-6. Boldface marks for a given indicator the solutions which are preferred. Italics mark the two solutions between which the highest amount of consistency can be obtained.
\\
\tablefoottext{a}{Measured [Fe/H] - mean cluster [Fe/H] from paper I.}\\
\tablefoottext{b}{Difference between the colour predicted using the spectroscopic $T_{\rm eff}$ with MARCS colour-temperature relations and the observed photometric colour dereddened assuming $E(B-V)=0.14$ and $E(V-I)/E(B-V)=1.244$.}\\
\tablefoottext{c}{Difference between the spectroscopic $\log g$ and $\log g$ calculated from the spectroscopic $T_{\rm eff}$ and the luminosity obtained using the observed $V$-magnitude, $(m-M)_V=13.51$, MARCS BC and different assumptions of the mass.}\\
\tablefoottext{d}{Mass for an assumed BSS star obtained from the extended ZAMS and evolutionary tracks in Fig.~\ref{fig:bhbvsbss}.}\\
\tablefoottext{e}{Mass for an assumed BHB star. A mass of about $0.52 M_{\sun}$ is implied when comparing the colour of 2-17 to the model ZAHB. However, since in this scenario 2-17 does not lie on the ZAHB in Fig.~\ref{fig:bhbvsbss}, a somewhat higher mass of $0.6 M_{\sun}$ was chosen to be sure that the mass is not underestimated.}\\
\tablefoottext{f}{The mass of 2-17 derived from the spectroscopic $T_{\rm eff}$ and $\log g$ in combination with luminosity obtained from the observed $V$-magnitude, $(m-M)_V=13.51$, and a MARCS BC.}
}
\end{center}
 
\end{table*}

\subsection{Blue straggler stars}
\label{sec:bss}

When comparing the positions of the blue stragglers in Fig.~\ref{fig:bhbvsbss} to the boundaries set by the ZAMS and the TAMS, we see that they are skewed towards the TAMS and slightly beyond. Following the results of \cite{Sills09} this is likely because the BSSs formed by joining the cores of two stars without much mixing with the material of the envelopes, and therefore the birth-line of the BSSs is quite far from a single star ZAMS because the material in the core has already been partly processed. For the same reason, the lifetime of a BSS is also much shorter than that of a normal star of the same mass. The presence of a rather high number of BSSs (even more evident in Fig. 1 of Platais et al. (2011) than in our Fig.~\ref{fig:bhbvsbss}) with short main sequence lifetimes suggests that BSSs are continuously being produced at a high rate and therefore BSSs should also be present in the later evolutionary stages within the cluster. Indeed, some of the stars marked with circles above the red clump in Fig.~\ref{fig:bhbvsbss}, which \cite{Platais11} suggested as binary HB stars could very well be evolved BSSs (E-BSS). Their positions and clustering above the RC in the CMD fit with this interpretation according to \cite{Sills09}. In addition, \cite{Miglio12} found a star in the RC which, despite being a cluster member \citep{Stello11}, does not show the seismic signature of a RC star, and has a seismic mass higher than the RC stars \citep{Corsaro12}. This star, marked with a square in Fig.~\ref{fig:bhbvsbss}, is likely to be an E-BSS in the RGB phase. Continued {\it Kepler} observations will provide further insights into these interesting stars.

\subsection{The white dwarf cooling sequence}
\label{sec:wd}

As an additional twist, we note that if many stars in NGC\,6791 have evolved through the BSS channel, as suggested, then the white dwarf (WD) cooling sequence should also harbour a significant fraction of E-BSSs. 

The WD cooling sequence of NGC\,6791 has become known in the literature for its peculiar double peak feature and the attempts to explain it (\citealt{Bedin05,Bedin08,Bedin08b}). \cite{Bedin08b} suggest that the bright peak is due to WD binaries, which would produce the observed magnitude-difference between the two peaks. However, we note that their simulated CMD produces the two peaks at the same colour, whereas the observations show the brighter peak to be bluer than the fainter (Fig. 2, panels b and c in \citealt{Bedin08b}). 

Therefore, we speculate that the E-BSS progenitors of some of the WDs we observe today might play a role in the appearance of the bright peak of the WD luminosity function of NGC\,6791. This could perhaps explain the colour-offset between the two peaks as a mass-effect; an E-BSS would be more massive and bluer than a normal star arriving on the WD cooling sequence at the same time. However, this remains speculation to encourage future investigations, as it calls for extensive and quantitative demonstrations which are beyond the scope of this paper. The binary WD scenario is also not ruled out, and it is likely that a combination of E-BSS and binary WDs are causing the complex WD cooling sequence of NGC\,6791.

\section{Summary, conclusions and outlook}
\label{sec:concl}

In this paper we presented an extensive comparison of stellar models to photometric and spectroscopic 
measurements of the eclipsing binaries V18 and V20, which are members 
of the old open cluster NGC\,6791.

By performing a combined stellar model comparison with {\it multiple}
eclipsing binaries in mass-radius and mass-$T_{\rm eff}$ diagrams and cluster CMDs we showed
that we can constrain stellar models better than ever before. This
allowed us to constrain the helium content, and thereby obtain a more
precise (and hopefully accurate) age of the cluster, although in a
model-dependent way, while demonstrating remaining age dependencies. 

Our model comparisons result in $Y=0.30\pm0.01$, resulting in $\Delta Y/\Delta Z \sim\, $1.4, assuming that such a relation exist and applies to NGC~6791. It therefore seems that the cluster proto-cloud was rather normal despite having super-solar metallicity at an old age.

NGC~6791 contains a large number of additional detached
eclipsing binaries (\citealt{Rucinski96, deMarchi07, Mochejska05}, Sandquist et al., in prep.). Accurate
measurements of these, which we have shown to be possible, would allow
even tighter model constraints by extending the mass range
over which the observed mass--radius relation must be reproduced. This
has the potential to strengthen the procedure followed in this paper. Furthermore,
NGC~6791 is in the Field-Of-View of the NASA {\it Kepler} mission \citep{Borucki10}, which will not
only allow many more detached eclipsing binaries to be found and measured, but also complementary model
constraints from asteroseismology of the giant stars in the cluster \citep{Basu11, Hekker11, Pablo11, Stello11, Corsaro12, Miglio12}. 
All our acceptable model fits give a mass on the RGB at the $V$-magnitude of the RC in the range $1.15\pm0.02 M_{\sun}$ which will be an important consistency check, along with the age and helium content, for future detailed asteroseismic modelling of the cluster giants.

Accurate surface gravities from both eclipsing binaries and asteroseismology can strongly aid future spectroscopic studies aimed at the abundance-related issues which we have shown to be the dominant contributor to the remaining age uncertainty of NGC\,6791.

As a by-product of our analysis we were able to correctly classify cluster BSSs and find indications of the presence of their evolved counterparts. The existence of E-BSS stars in the RC phase can be confirmed with future {\it Kepler} data.

Multiple detached eclipsing binaries have also been confirmed in a
number of other old open clusters (\citealt{Brogaard10,Talamantes10}, Sandquist et al., in prep.) including NGC6819, which is also in the {\it Kepler} Field-Of-View.
Extending this kind of analysis to these clusters will allow more
reliable cluster ages to be determined and more aspects of stellar
models to be tested, when stellar models have to reproduce observations
in clusters with a range in age and metallicity. It is our hope that this will also help disentangle some of the issues related to colour-temperature transformations and/or the model $T_{\rm eff}$ scale.

By extending studies also to the globular clusters, detached eclipsing binaries could
ultimately provide the strongest constraints on their ages and helium
contents.

\begin{acknowledgements}
This paper is dedicated to J. V. Clausen who devoted his career to eclipsing binary star research and was an inspiration to many of us.

We thank R. Peterson and E. Green for kindly proving their spectrum of the star 2-17 and K. Cudworth for providing his preliminary proper motion membership data of cluster stars.

We also thank an anonymous referee for useful comments, which helped improve the paper.

KB acknowledges financial support from the Carlsberg Foundation.
DAV acknowledges financial support from the Natural Science and Engineering Reasearch Council of Canada.
DS acknowledges support from the Australian Research Council.

The following internet-based resources were used in research for
this paper: the NASA Astrophysics Data System; the SIMBAD database
and the ViziR service operated by CDS, Strasbourg, France; the
ar$\chi$iv scientific paper preprint service operated by Cornell University.
\end{acknowledgements}

\newpage

\bibliographystyle{aa}
\bibliography{11} 




\end{document}